\documentclass[12pt,preprint]{aastex}
\usepackage{mkfig}
\begin{document}

\title{\bf The Coronal Abundances of Mid-F Dwarfs}

\author{Brian E. Wood, J. Martin Laming}

\affil{Naval Research Laboratory, Space Science Division,
  Washington, DC 20375}


\begin{abstract}

     A {\em Chandra} spectrum of the moderately active nearby F6~V star
$\pi^3$~Ori is used to study the coronal properties of mid-F dwarfs.
We find that $\pi^3$~Ori's coronal emission measure distribution is
very similar to those of moderately active G and K dwarfs, with an
emission measure peak near $\log T=6.6$ seeming to be ubiquitous for
such stars.  In contrast to coronal temperature, coronal abundances
are known to depend on spectral type for main sequence stars.
Based on this previously known relation, we expected $\pi^3$~Ori's
corona to exhibit an extremely strong ``FIP effect,'' a phenomenon
first identified on the Sun where elements with low ``First Ionization
Potential'' (FIP) are enhanced in the corona.  We instead find that
$\pi^3$~Ori's corona exhibits a FIP effect essentially identical to
that of the Sun and other early G dwarfs, perhaps indicating that the
increase in FIP bias towards earlier spectral types stops or at
least slows for F stars.  We find that $\pi^3$~Ori's coronal
characteristics are significantly different from two previously
studied mid-F stars, Procyon (F5~IV-V) and $\tau$~Boo (F7~V).  We
believe $\pi^3$~Ori is more representative of the coronal
characteristics of mid-F dwarfs, with Procyon being different
because of luminosity class, and $\tau$~Boo being different because
of the effects of one of two close companions, one
stellar ($\tau$~Boo~B:  M2~V) and one planetary.

\end{abstract}


\section{INTRODUCTION}

     Although details about stellar magnetic field generation
remain uncertain, it is accepted that coronal magnetic fields are
generated by a magnetic dynamo operating within the convection zone.
For main sequence stars, the convection zone narrows towards earlier
spectral types, disappearing entirely around a spectral type of A5~V.
At this point, coronae disappear as well \citep{js85},
emphasizing the central role that the convection zone dynamo plays in
generating stellar coronae.

     One general empirical goal of stellar coronal X-ray observations
is to study how coronae change character as the convection zone narrows.
Any clear correlation between coronal properties and spectral type
tells us something useful about how the dynamo and the fields it
generates change with spectral type, thereby providing potentially
crucial information for theoretical models of how stellar magnetic
dynamos operate in general.

     Coronal abundances represent one coronal property that appears to
be tightly correlated with spectral type, at least for main sequence
stars.  On the Sun, it has long been known that coronal abundances are
different from photospheric abundances, elements with low First
Ionization Potential (FIP) having enhanced abundances in the solar
corona and solar wind \citep{rvs95,uf00}.
Analyses of X-ray and EUV spectra have found a similar ``FIP
effect'' for some stars of low to moderate activity
\citep{jml96,jjd97,jml99,at05}.
However, for many other stars, particularly extremely
active ones, the FIP effect is either absent, or sometimes an inverse
FIP effect is observed, where low-FIP elements have coronal abundances
that are {\em depleted} relative to the high-FIP elements
\citep{ma01,ma03,acb01,mg01,dph01,dph03,jsf03,jsf09,bb05,cl08}.
Currently the only detailed
theoretical model that is capable of explaining both solar-like FIP
effects and inverse FIP effects is one that attributes the element
fractionation to ponderomotive forces induced by Alfv\'{e}n and other
MHD waves passing through or reflecting from the chromosphere
\citep{jml04,jml09,jml12,bew12}.

     Initially, it appeared that coronal abundances were primarily
dependent on activity, as the most active stars clearly tend to have
inverse FIP effects instead of a solar-like FIP effect.  This activity
dependence of abundances for active stars is illustrated and discussed
in detail by \citet[][see Figure 37]{mg04}.  The abundance variations
are not simply a uniform variation of low-FIP elements relative to
high-FIP ones.  The high-FIP elements appear to fractionate with
respect to each other as well, as illustrated by measurements of
Ne/O.  The vast majority of spectroscopically observed stellar coronae
exhibit high ratios of ${\rm Ne/O}\approx 0.4$ \citep{jjd05}, but
relatively inactive stars (e.g., the Sun, $\alpha$~Cen~AB, and Procyon)
show ${\rm Ne/O}\approx 0.2$ \citep{jr08}.

     As more abundance measurements have been made, it has
become apparent that abundances are also highly dependent on spectral
type \citep{mg07a,mg07b,at07}.  In order to explore this further we
have experimented with a high-FIP/low-FIP abundance ratio, $F_{bias}$,
which quantifies a corona's FIP bias as the average abundance of four
high-FIP elements (C, N, O, and Ne) relative to Fe; normalized to
photospheric abundance ratios.  For normal main sequence stars with
coronal X-ray luminosities of $\log L_X<29.1$ (in ergs~s$^{-1}$),
there is a surprisingly tight spectral type dependence for
$F_{bias}$ \citep{bew10,bew12}.  In this correlation, early G stars
all have a solar-like FIP effect, which decreases towards late G and
early K stars, reaching no FIP effect at all at a spectral type of
K5~V.  Later than K5~V, inverse FIP effects are observed, with the
magnitude of the effect increasing into the M dwarfs.  We will refer
to this as the ``FIP-Bias/Spectral-Type'' (FBST) relation.  At least
for stars with $\log L_X<29.1$, any activity dependence for $F_{bias}$
is lost in the scatter of the measurements.

     The FBST relation would predict that the solar-like FIP effect
should become even stronger as you move from solar-like early G stars
to F stars.  More precisely, instead of a factor of 4 enhancement of
low FIP elements in the corona as for the Sun \citep{uf00},
the FBST relation would predict a factor of 6 enhancement for
mid-F stars \citep{bew10}.  However, observations of two mid-F
dwarfs seem to contradict this expectation.  The first of these is
Procyon (F5~IV-V), the nearest and most frequently observed F star,
which has been observed spectroscopically by EUVE, {\em Chandra}, and
XMM.  All of these spectra demonstrate that Procyon's corona exhibits
no significant FIP bias at all.  Its coronal abundances are consistent
with its photospheric abundances \citep{jjd95,ajjr02}.
More recently, using XMM data \citet{am11} have
also found little difference in the abundances of low-FIP and high-FIP
elements in the corona of $\tau$~Boo (F7~V), although they suggest
that {\em both} are depleted in the corona by similar amounts.

     The Procyon and $\tau$~Boo measurements would seem to suggest
that the FBST relation falls apart for the F dwarfs, possibly an
effect of the thinning convection zones of such stars.  This would be
a very interesting result, but there are potential issues with both
Procyon and $\tau$~Boo regarding whether either can be considered
representative of mid-F dwarfs.  For Procyon, the problem is simply
that it is not a pure main sequence star, as its ``IV-V'' luminosity
class suggests.  Its radius of 2.05 R$_{\odot}$ is certainly larger
than that of a true mid-F main sequence star \citep{pk04}.

     For $\tau$~Boo, the problem is the presence of two very different
companions, one stellar and one planetary.  The stellar companion,
$\tau$~Boo~B, is an M2~V star that is unresolved from the primary in
the XMM observation, which could in principle be contributing X-ray
photons to the spectrum.  Given that M dwarfs are expected to have
inverse FIP effects \citep{bew12}, contamination from
$\tau$~Boo~B would be expected to weaken any FIP effect in the
combined X-ray spectrum.  As for the planetary companion, it is a very
well-studied close-in giant planet, $\tau$~Boo~b, with an orbital
period of only 3.31 days.  There is already evidence that the presence
of this massive planet is producing coronal activity on the star
\citep{es08,gahw08}.  Is it possible that the
planet could somehow be suppressing the expected strong coronal FIP
effect as well?  This is an exciting possibility, which
makes it all the more important to know what the corona of a truly
normal mid-F dwarf looks like.

     To that end, we here present an analysis of recent {\em Chandra}
observations of the nearby F6~V star, $\pi^3$~Ori, which lies at a
distance of 8.1~pc \citep{macp97}.  The star has a rapid rotation rate
of $v\sin i=17.3$ km~s$^{-1}$, corresponding to $P_{rot}/\sin i=4.1$
days \citep{mave12}.  Not surprisingly, given the known connection
between stellar rotation and activity, $\pi^3$~Ori also has a
relatively high X-ray luminosity of $\log L_X=28.96$ \citep{js04},
just barely within the $\log L_X<29.1$ activity regime mentioned above
where the FBST relation remains tight.  This is in fact one of the ten
most luminous coronal X-ray sources within 10 pc, so it is curious
that $\pi^3$~Ori has never before been observed spectroscopically in
X-rays.  The $\pi^3$~Ori spectrum promises to resolve the issue of
what the coronal characteristics of moderately active mid-F main
sequence stars really are.

\section{DATA ANALYSIS}

\subsection{Observations of $\pi^3$~Ori}

     The {\em Chandra} X-ray telescope observed $\pi^3$~Ori twice
in late 2010, the first time on 2010~November~11 for 57.8~ksec, and a
second time on 2010~November~23 for 20.1~ksec.  The observations were
performed with {\em Chandra}'s LETGS configuration, combining the Low
Energy Transmission Grating (LETG) and the component of the High
Resolution Camera detector used for spectroscopy (HRC-S).

     The data were processed using version 4.3 of the {\em Chandra}
team's CIAO software.  The initial HRC-S image produced by the
data processing consists of a zeroth-order image of the star, and
an X-ray spectrum dispersed on both sides of the image, i.e., a
plus and minus order spectrum.  Further processing extracts
one-dimensional spectra from the image, and then the plus and minus
order spectra are coadded into a final spectrum covering the full
$5-175$~\AA\ wavelength region of LETGS.

     Photons in the time-tagged zeroth-order image are useful
for assessing source variability.  We find no significant
variability at all for $\pi^3$~Ori, and the X-ray fluxes appear to
be identical at the two observing times, so we simply coadd the
spectra taken at the two times into a final spectrum representing
a combined exposure time of 77.9~ksec.  This spectrum is shown in
Figure~1.

\subsection{Line Identification and Measurement}

     We use version 6.1 of the CHIANTI atomic database to identify
emission lines in our spectrum \citep{kpd97,kpd09}.  The
identified lines are listed in Table~1, and identifications are also
provided in Figure~1.  Counts for the detected lines are measured by
direct integration from the spectrum, and these raw counts are then
converted to photon fluxes.  Both the counts and fluxes of the lines
are provided in Table~1, together with 1$\sigma$ uncertainties.  The
third column of the table lists line formation temperatures, based on
ionization equilibrium computations from \citet{ma85}.

     There are many lines in Table~1 with large quoted uncertainties
that represent questionable detections.  The emission measure (EM)
analysis described in section 2.3 is used to confirm {\em a
  posteriori} that these marginal detections are at least plausible.
There are many line blends in Table~1.  The EM analysis and the line
strengths in the CHIANTI database are used to assess what lines are
contributing significant flux to each emission feature.  The line
blends involving different species are ultimately ignored in the final
EM analysis.

     There are many density sensitive He-like triplets in the $\pi^3$~Ori
spectrum \citep{jun02}, but only the O~VII triplet near
21.8~\AA\ is sufficiently free of blends and with high enough
signal-to-noise (S/N) for a decent density measurement.
Unfortunately, even for O~VII we find we can only quote an upper limit
for the electron density.  Based on the collisional equilibrium models
of \citet{dp01}, the $\lambda22.101/\lambda21.807$ line flux
ratio corresponds to an electron density (in cm$^{-3}$) of $\log
n_e<10.70$.  Similarly low coronal densities have been found for many
moderately active G and K dwarfs in the past \citep{bew06,bew10},
so there is no evidence that our moderately active F dwarf is
any different from later type stars in terms of coronal density.

\subsection{Emission Measure Analysis}

     Inferring coronal abundances and temperature distributions
from line flux measurements requires an EM analysis, which is a
nontrivial inversion problem.  For this calculation, we use version
2.6 of the PINTofALE software package \citep{vk98,vk00},
which was developed to solve the EM inversion problem using a Markov
chain Monte Carlo approach.  The analysis also makes use of CHIANTI
line emissivities \citep{kpd97,kpd09}.  Our EM analysis closely
mirrors previous work, as described in particular detail by \citet{bew06},
so we refer the reader to that paper for details about
this computation.

     The analysis considers the effects of interstellar absorption
on the observed line fluxes.  There is no measured H~I column density
towards $\pi^3$~Ori itself, but there is a measurement of $\log
N_H=17.93$ (in cm$^{-2}$) towards $\chi^1$~Ori \citep{bew05},
which is roughly at the same distance and in the same direction as
$\pi^3$~Ori, so this is the value we assume.  We note, however, that
this value is low enough that interstellar absorption does not have a
large effect on line fluxes.

     The line fluxes by themselves allow the computation of
the shape of the emission measure distribution, and the computation of
relative coronal abundances.  For the relative abundances, the natural
reference element to use is Fe, considering the large number of Fe
lines available in the LETGS spectrum.  In the initial EM analysis we
simply assume a solar photospheric abundance for our reference
element, $\log {\rm [Fe/H]}=-4.50$ \citep{ma09}.  Table~2 lists the coronal
abundances relative to Fe computed in the EM analysis, and Figure~2
shows the derived emission measure distribution.  The 90\% confidence
intervals shown for the emission measures and abundances in Figure~2
and Table~2, respectively, are indicative of the uncertainties in
solving the inversion problem, based on the magnitude of the
uncertainties in the measured line fluxes.

     Properly normalizing the EM distribution requires the measurement
of an absolute coronal [Fe/H] ratio, which also allows the relative
abundances in Table~2 to be converted to absolute abundances.
Measurement of [Fe/H] requires an assessment of the line-to-continuum
ratio in the LETGS spectrum, which can be done after the initial EM
analysis is completed.  Emission measures scale inversely with the
assumed [Fe/H], since higher metal abundances mean less emission
measure is required to account for the line fluxes.  Thus, increasing
[Fe/H] by a factor of two from $\log {\rm [Fe/H]}=-4.50$ to
$\log {\rm [Fe/H]}=-4.20$, for example, uniformly lowers the EM
distribution by a factor of two.

     Synthetic line-plus-continuum spectra can be computed from the
EM distribution for various values of [Fe/H] and compared with the data,
in order to see which leads to the best match to the observed continuum
level.  This is illustrated in Figure~3, which shows the best fit for
$\log {\rm [Fe/H]}=-4.32$.  This synthetic spectrum is also compared
with the data in Figure~1.  In LETGS data, higher order spectra
are superimposed onto the first order spectrum, so the synthetic
spectrum accounts for the higher order spectra as well, specifically
orders 2--5.  The higher order contributions are explicitly plotted
in both Figures~1 and 3.  The EM distribution in Figure~2 is the final
normalized distribution assuming $\log {\rm [Fe/H]}=-4.32$.

\subsection{The Shape of the Emission Measure Distribution}

     The most striking feature of the $\pi^3$~Ori EM distribution
in Figure~2 is a sharp peak at $\log T=6.6$.  This peak actually seems
to be very common for moderately active stars like $\pi^3$~Ori.  Many
examples are provided by \citet{bew06,bew10}, but the best of
these are $\xi$~Boo~A (G8~V) and $\epsilon$~Eri (K2~V).  The EM
distributions for these 2 stars are also shown in Figure~2.  The
$\xi$~Boo~A EM distribution is a particularly good match for that of
$\pi^3$~Ori.  This is perhaps not surprising considering that
$\xi$~Boo~A and $\pi^3$~Ori have similar X-ray luminosities of $\log
L_X=28.86$ and $\log L_X=28.96$, respectively \citep{js04},
although the spectral types of the two stars are quite
different.  Thus, there is no evidence that the
thinner convection zone of $\pi^3$~Ori has led to coronal temperatures
any different from those of similarly active later type stars.

     Interestingly enough, the $\log T=6.6$ peaks are also commonly
observed for solar active regions \citep{hpw11,hpw12}.  In
Figure~2, the three stellar EM distributions are compared with a solar
one, which represents the average of 15 active region EM distributions
from \citet{hpw12}, who emphasized the prevalence of
$\log T=6.6$ EM peaks on the Sun.  The shape of the solar active
region EM agrees beautifully with those of $\pi^3$~Ori and $\xi$~Boo~A
for $\log T<6.7$, although the three stellar EM distributions in
Figure~2 have much higher EM at $\log T>6.7$.  \citet{jjd00}
have previously noted the similarity between the EM distributions of
solar active regions and those of $\xi$~Boo~A and $\epsilon$~Eri, and
have suggested that it implies that moderately active stars like these
are covered almost completely with solar-like active regions.

     Is it possible that these seemingly ubiquitous $\log T=6.6$ peaks
are somehow a systematic artifact of the PINTofALE computations that
produced them?  Not only was PINTofALE's Markov chain Monte Carlo
approach used to compute the stellar EM distributions in Figure~2, but
it was also used by \citet{hpw11,hpw12} to compute the solar
ones.  Note, however, that \citet{hpw11,hpw12} analyzed a completely
different spectral region and a different set of emission lines than
those used in the stellar analyses, suggesting that the $\log T=6.6$
peaks are not just a product of working with a particular set of
lines.  We explore this issue further using simulations.  These
simulations involve taking a known EM distribution, computing line
fluxes from that distribution for the set of lines in Table~2, and
then plugging those line fluxes back into PINTofALE to see if the EM
inversion procedure can recover the original EM distribution.
Uncertainties in the line fluxes are assumed to be the same as the
relative uncertainties measured for the $\pi^3$~Ori lines in Table~1.
Some results of these simulations are shown in Figure~4.

     The dotted line in Figure~4's ``Simulation~1'' is simply the
$\pi^3$~Ori distribution from Figure~2.  The solid line shows the
result when that is used as the input EM distribution for the
simulation.  The computed EM distribution agrees encouragingly well
with the original distribution, being within the displayed error bars
for 15 of the 17 temperature bins.  Most importantly, the simulation
demonstrates PINTofALE's ability to recover the sharp $\log T=6.6$ EM
peak.  In Simulations~2 and 3 we simply shift the peak to $\log T=6.5$
and $\log T=6.7$, respectively, and in Simulation~4 we remove the peak
entirely, assuming equal EM values in the $\log T=6.5-6.7$ range.
Although the reconstructed EM values are still generally within error
bars, we do note that for Simulations~2--4 the EM is systematically
overpredicted at $\log T=6.6$ and underpredicted at $\log T=6.5$ and
$\log T=6.7$.

     We conclude, therefore, that there may indeed be a modest
tendency for PINTofALE to overemphasize the height and narrowness of
the $\log T=6.6$ peaks, though the magnitude of this effect is not
necessarily inconsistent with the inferred 90\% confidence intervals.
We can only speculate as to the source of this tendency.  It does so
happen that $\log T=6.6$ is the central formation temperature of the
strongest lines in the LETGS spectrum, the Fe~XVII lines between
15--17~\AA, so perhaps the need to fit those line fluxes particularly
well leads to the modest overemphasis of EM at $\log T=6.6$, to
the detriment of EM at $\log T=6.5$ and $\log T=6.7$.  In any case,
despite evidence for this small systematic error, the simulations
provide support for the existence of a distinct maximum of EM near
$\log T=6.6$ for moderately active stars and solar active regions.

     The question then becomes, why is $\log T=6.6$ such a popular
temperature for both a variety of solar active regions, and for
moderately active stars of various spectral types?  Is there something
special about this particular temperature?  One notable thing about it
is that there is a local minimum in the radiative cooling curve near
$\log T=6.6$ \citep[e.g.,][]{jc08,jd11}.  The
cooling curve relates how emissivity varies with temperature.  Plasma
is stable against radiatively driven thermal instabilities where the
cooling curve has a shallower slope than the temperature dependence of
the heating function.  The cooling curve has a steep negative slope
for a variety of abundance sets in the range $\log T=6.1-6.5$, which
could imply thermal instability unless the heating curve has an even
steeper negative slope at those temperatures.  Thus, $\log T=6.6$ is
possibly the coolest stable temperature above $\log T=6.1$.  Plasma
initially between these two temperatures will most likely either
heat up or cool down until a stable temperature is reached.

     Perhaps the ubiquity of the $\log T=6.6$ peak implies something
about the commonality of active region loops on the Sun and solar-like
main sequence stars, consistent with the idea that coronal emission
from moderately active stars like $\pi^3$~Ori and $\xi$~Boo~A is
dominated by solar-like active regions that largely cover the stellar
surface \citep{jjd00}.  On the Sun, such loops are often found
to have densities of $n_e \sim 10^{10}$ cm$^{-3}$ \citep[e.g.,][]{arw11}.
Such densities are consistent with those commonly
measured in stellar coronae as well, at least at temperatures of $\log
T<6.7$ \citep{jun02,bew06,bew10}.  An explanation
in terms of thermal stability presupposes that coronal plasma exists
in structures (loops) that can support a quasi-steady state.  No such
explanation would be possible in a corona heated and supplied by more
dynamic phenomena such as Type II spicules
\citep[e.g.,][]{bdp11,swm11,jms11,pgj12}.  In simple static loop models
there is a relation between a coronal loop's length and its apex
temperature and density \citep{rr78,jfv79,cj80,fr10}.
Thus, if coronal loops in various solar and stellar contexts tend to
have similar densities and lengths, then presumably they would have
similar temperatures as well.  This presumes, of course, that these
simple static models apply, which is
questionable \citep[e.g.,][]{ddl99,jts09}.

\subsection{The Coronal Abundances of $\pi^3$~Ori}

     As discussed in section 2.3, the EM analysis provides coronal
abundances measured relative to Fe for many elements.  All of these
measurements, in the form $\log{\rm [X/Fe]}$, are listed in the second
column of Table~2.  We are interested in comparing coronal abundances
with those of the photosphere.  Thus, in the third column of Table~2,
we list stellar photospheric abundance ratios relative to Fe, in the
form $\log{\rm [X/Fe]}_*$.  The photospheric abundances of \citet{cap04}
are measured using line-by-line comparisons of
solar and stellar spectral features, so the abundance measurements are
intrinsically relative to solar abundances.  Thus, computing the
stellar photospheric abundance ratios in Table~2 requires the
assumption of solar abundances.  The assumed solar abundances of
\citet{ma09} are listed in the last column of Table~2.  For
N and S, there are no stellar photospheric measurements available.  We
assume these elements behave like high-FIP elements, such as O.  Thus,
we assume $\log{\rm [N_*/N_{\odot}]}\equiv \log{\rm
[S_*/S_{\odot}]}\equiv \log{\rm [O_*/O_{\odot}]}=0.16$.

     Neon is a special case.  The fundamental problem with Ne is
that even for the Sun there is no real photospheric abundance
measurement, because there are no solar photospheric Ne absorption
lines.  Reference solar abundances for Ne, such as the one listed in
Table~2, are in reality based on measurements from the solar corona
and transition region \citep[e.g.,][]{jts05}.  We know very well
that abundances can be different from photospheric in the corona, but
not necessarily in the transition region, where \citet{pry05a,pry05b}
finds a low Ne/O abundance, and no evidence for a FIP effect in other
elements.  Nevertheless, we do know that high-FIP elements can be
fractionated in the solar atmosphere, as He is depleted by a factor of
2 in the slow solar wind \citep{rvs95}.  This depletion
of He (and also possibly of Ne) is also a feature of models for the
solar FIP effect based on the ponderomotive force \citep{jml09,jml12,cer12}.

     Complicating the issue further is the knowledge that moderately
active and very active stars seem to univerally have coronal Ne/O
ratios much higher than that of the Sun and other similarly inactive
stars \citep{jjd05,bew06,jr08},
raising the possibility that these stellar coronal Ne
abundances may be a better measure of the true cosmic Ne abundance
than the solar coronal Ne abundance usually referenced as such.  We
here follow the same convention as in our past papers and assume that the
average stellar coronal ${\rm Ne/O}=0.41$ abundance ratio of \citet{jjd05}
is indicative of the true cosmic abundance of Ne, and
therefore applies to solar and stellar photospheres as well, including
for $\pi^3$~Ori.  Thus, the $\log {\rm [Ne/Fe]}_*$ ratio in Table~2 is
derived assuming ${\rm Ne/O}=0.41$, instead of the much lower solar
ratio of ${\rm Ne/O}=0.17$.  The validity of this assumption is very
much debatable, but it does lead to coronal Ne abundances in better
agreement with those of other high-FIP elements (see Figure~5).

     After subtracting the logarithmic photospheric abundance ratios,
the logarithmic coronal abundance ratios in Table~2 are in Figure~5a
plotted versus FIP.  The low-FIP elements are all roughly consistent
with the reference low-FIP element, Fe.  The high-FIP element
abundances are lower by about $-0.6$ dex, consistent with a FIP effect
nearly identical to that of the Sun.  The average value of the four
high-FIP elements in Figure~5a is plotted explicitly as a horizontal
dot-dashed line in the figure.  This is the ``FIP bias'' quantity,
$F_{bias}$, that we have used in the past to reduce coronal abundances
to a single number, which is then used to study the FBST relation
described in section~1 \citep{bew10,bew12}.  For $\pi^3$~Ori,
$F_{bias}=-0.55$.

     The $F_{bias}$ quantity has no clear variation with activity
for main sequence stars with $\log L_X<29.1$ \citep[see Figure 4
in][]{bew12}, but it is worth noting that of the four high-FIP
elements that go into computing $F_{bias}$ (C, N, O, and Ne), Ne is
once again an exception, as we know that the coronal Ne/O ratio
does vary within the $\log L_X<29.1$ sample, suggesting
activity-dependent fractionation of Ne even for less active stars
\citep{jr08}.  As alluded to above, the vast majority of
spectroscopically observed $\log L_X<29.1$ stars, including
$\pi^3$~Ori, have ${\rm Ne/O}\approx 0.4$.  Only a few truly inactive
stars (the Sun and $\alpha$~Cen~AB) have ${\rm Ne/O}\approx 0.2$
\citep{bew12}.

     There are a couple characteristics of the low-FIP elements
that are worthy of note.  \citet{am11} found that the coronal
Ni abundance of the mid-F dwarf $\tau$~Boo was anomalously high
compared to other low-FIP elements, being a factor of $2-3$ higher.
We do not see this behavior for our mid-F dwarf, $\pi^3$~Ori.  The Ni
abundance may be a little high in Figure~5, but it is consistent with
the other low-FIP elements within the error bars.  As for Mg and Si,
in the past we had noted a tendency for Mg abundances to be higher
than Si abundances in the coronae of many moderately active stars
\citep{bew10}.  This is not the case for $\pi^3$~Ori, though,
which if anything shows the opposite behavior, consistent with what
\citet{am11} found for $\tau$~Boo.

     Figure~5a shows the coronal abundance measurements relative
to Fe.  But using the $\log {\rm [Fe/H]}=-4.32$ measurement from the
line-to-continuum analysis, we can compute absolute abundances as
well, and in Figure~5b the absolute abundances are plotted, relative to
stellar photospheric abundances as before.  This figure casts doubt on
whether the coronal abundance fractionation pattern of $\pi^3$~Ori is
truly solar-like or not.

     Based on Figure~5a, the $\pi^3$~Ori coronal abundances were
described above as very solar-like, because the $F_{bias}=-0.55$
measurement implies a factor of 4 underabundance of high-FIP elements
relative to low-FIP elements in the corona, similar to the situation
in the solar corona.  But in the solar corona this is really due to an
enhancement of low-FIP elements, as opposed to a depletion of high-FIP
elements, meaning that for the Sun the high-FIP elements would be near
0.0 in Figure~5b and the low-FIP elements would be near 0.6.  In
contrast, Figure~5b would imply that for $\pi^3$~Ori, the
$F_{bias}=-0.55$ result is due to fractionation of {\em both} low-FIP
and high-FIP elements, and is in fact mostly due to a depletion of
high-FIP elements.

     This is actually consistent with what we have found before for
other moderately active stars \citep{bew06}.  The question is
whether this means that the fractionation behavior of these stars
really is significantly different from the Sun, or whether there is
a systematic error in the line-to-continuum analysis that leads to
systematic underestimates of $\log {\rm [Fe/H]}$.  The history
of solar coronal abundance measurements provides reason for caution with
regards to the accuracy of the line-to-continuum analysis, as
many such solar analyses have found the data more consistent with
a high-FIP depletion than a low-FIP enhancement \citep{njv81,af95},
contradicting other evidence, including direct
solar wind measurements, that suggest low-FIP enhancement is
the primary effect \citep{rvs95,uf00}.

     One concern is the possible presence of weak metal lines not
listed in atomic line databases.  These lines could collectively
produce a significant pseudo-continuum in X-ray spectra, leading to
overestimates of the strength of the real continuum, corresponding to
an underestimate of $\log {\rm [Fe/H]}$.  There is hope that modern
databases such as CHIANTI with more extensive line lists have greatly
reduced this problem, but it is difficult to quantify what the
systematic uncertainties really are.  Another potential issue is the
effect of non-Maxwellian distributions on the line-to-continuum
analysis.  Non-Maxwellian distributions will affect the analysis of
both the line and continuum fluxes, but in different ways \citep{jd11,jd12}.

     All these issues provide justification for relying primarily on
relative, rather than absolute abundances.  The latter introduces the
uncertain systematic errors involved in the line-to-continuum
analysis, while the former does not.  Nevertheless, when using
relative abundances, and quantities derived from them such as
$F_{bias}$, it is important to keep in mind that the issue of absolute
abundances has been left unresolved.  A negative $F_{bias}$ value, for
example, could indicate either a low-FIP enhancement or a high-FIP
depletion, and two stars with identical $F_{bias}$ values could in
principle be different in that respect.  That being said, it is hard
to imagine how the rather tight FBST relation shown in Figure~6 (and
discussed in detail in the next section) could exist without some
consistency regarding how $F_{bias}$ relates to absolute abundances.
For example, the FBST relation suggests M4~V stars all have
$F_{bias}\approx 0.45$ \citep{bew12}.  It is hard to imagine why
these stars would have different degrees of low-FIP depletions and/or
high-FIP enhancements, but always calibrated so that $F_{bias}\approx
0.45$.

\section{REASSESSING THE CORONAL ABUNDANCE PROBLEM}

\subsection{$\pi^3$~Ori and the FBST Relation}

     As described in section~1, one of the main goals of the
$\pi^3$~Ori analysis is to see whether the FBST relation extends to
F spectral types.  The answer is provided in Figure~6.  A full list
of the stars plotted in the figure is provided by \citet{bew12},
but we now add the $\pi^3$~Ori data point.

     Based on the G, K, and M stars in the figure,
$\pi^3$~Ori would be expected to have $F_{bias}\approx -0.8$,
potentially the strongest solar-like FIP effect ever observed.
Instead it has $F_{bias}=-0.55$, comparable to the Sun and other early
G stars.  There are three possible interpretations for the
higher-than-expected $F_{bias}$ measurement for $\pi^3$~Ori:
\begin{enumerate}
\item Simply taken at face value, the $\pi^3$~Ori measurement
  indicates that the FBST relation flattens towards spectral types
  earlier than G.  Observations of early F dwarfs would be very useful
  to verify this flattening.  If real, the flattening of the FBST
  relation could be interpreted as an interesting manifestation of the
  narrowing convection zone.
\item Although somewhat higher than expected, it could be argued that
  the $\pi^3$~Ori measurement is still close enough to the expected
  $F_{bias}\approx -0.8$ value to be considered consistent with a
  near-linear FBST relation, considering the amount of scatter that
  one sees in Figure~6.
\item Perhaps $\pi^3$~Ori is actually too active to precisely follow
  the FBST relation.  It is known that stars with $\log L_X>29.1$ lie
  significantly above the FBST relation \citep{bew12}.  With
  $\log L_X=28.96$ \citep{js04}, $\pi^3$~Ori is close to this limit.
  If the $L_X$ limit is actually spectral type dependent, and lower
  for F dwarfs, $\pi^3$~Ori would be expected to lie somewhat above
  the FBST curve.
\end{enumerate}

     In section~1, there were two F dwarfs discussed that are
known from previously published work to be inconsistent with the FBST
relation:  Procyon (F5~IV-V) and $\tau$~Boo (F7~V).  Based on
measurements from \citet{ajjr02} and \citet{am11}, we
compute values of $F_{bias}=0.12$ and $F_{bias}=-0.17$ for Procyon and
$\tau$~Boo, respectively.  These values are plotted in Figure~6,
explicitly showing just how inconsistent these two stars are with both
$\pi^3$~Ori and the FBST relation.

     As for Procyon, the problem is that it is not a pure
main sequence star, as its ``IV-V'' luminosity class and large
2.05~R$_{\odot}$ radius demonstrate \citep{pk04}.  Not only
does Procyon have coronal abundances very different from $\pi^3$~Ori,
but it has a much cooler corona as well, with an EM peak near $\log
T=6.2$ and no significant EM above $\log T=6.4$ \citep{ajjr02}.
Cool coronae are associated with less active stars, but
Procyon's X-ray luminosity is $\log L_X=28.51$, only a factor
of 3 lower than $\pi^3$~Ori, and comparable to many of the stars in
the Figure~6 sample with hotter coronae.  For example, $\epsilon$~Eri
has a coronal X-ray luminosity of $\log L_X=28.33$ \citep{js04},
lower than Procyon, but it has an EM distribution similar to
$\pi^3$~Ori (see Figure~2).  We believe that Procyon's inconsistency
with $\pi^3$~Ori emphasizes just how sensitive coronal properties can
be to luminosity class.  Procyon simply cannot be considered to be a
main sequence star, and it is dramatically inconsistent with the FBST
relation as a consequence.

     As for $\tau$~Boo, the inconsistency of $\tau$~Boo with
$\pi^3$~Ori strengthens the argument made in section~1 that
$\tau$~Boo's high $F_{bias}$ value is due to one of its two
companions.  Either the M2~V companion, $\tau$~Boo~B, is contributing
substantial flux to the XMM exposure, leading to erroneous abundance
measurements, or the close-in giant planet, $\tau$~Boo~b, is somehow
affecting the coronal abundances.  The latter possibility is easily
the more interesting of the two, as it would add a new twist to the
current hot topic of star-planet interaction.

\subsection{What Determines Coronal Abundances?}

     At this point it is worthwhile to summarize what we know
about how coronal abundances depend on stellar properties.
The stellar properties that we know affect $F_{bias}$ are:
\begin{enumerate}
\item Spectral Type.  This is the FBST relation illustrated in
  Figure~6.  There are numerous basic stellar, photospheric,
  and convection zone properties that are known to be tightly
  correlated with spectral type for main sequence stars (stellar
  radius, temperature, mass, surface gravity, convection zone depth,
  etc.).  The existence of the FBST relation demonstrates
  that somehow one or more of these characteristics also determines
  coronal abundances.  It is worth noting that a spectral type
  dependence is also known to exist for T~Tauri stars and very
  active main sequence stars \citep{mg07b,bew12}.
\item Activity and/or Rapid Rotation.  The FBST relation seems to
  work very well for the vast majority of main sequence stars.
  However, there is clearly an activity dependence as well.  This is
  most evident for extremely active stars with $\log L_X>29.1$, which
  all lie significantly above the FBST relation in Figure~6
  \citep{bew12}.  This activity dependence is also clearly seen in
  Figure~37 of \citet{mg04}.  An indication that activity dependence
  of coronal abundances extends to less active stars is the difference
  in Ne/O seen between moderately active stars and truly inactive
  stars like the Sun \citep{jr08}.  For the activity extremes, it is
  possible that some aspect of high activity (e.g., high magnetic
  field strength) is affecting coronal abundances.  But it is also
  possible that the underlying cause of the extreme activity, rapid
  rotation, is the true determining factor.  Perhaps very rapid
  rotation affects photospheric and/or convection zone
  characteristics, thereby allowing coronal abundances to be altered
  in the same manner as spectral type affects abundances in the FBST
  relation.
\item Luminosity Class.  The best example of this is the
  $\pi^3$~Ori/Procyon dichotomy discussed in section 3.1.  There are
  evolved stars other than Procyon that would lie above the FBST
  relation in Figure~6, but most are part of active binaries
  \citep{acb01,dph01,ma01,ma03}.  For such stars, it is not clear
  whether the discrepancy from the FBST relation is truly due to
  luminosity class, or whether it is instead an effect of high
  activity or in some cases particularly close binarity.
\end{enumerate}

     Is there a common theme that can be found among these abundance
determinants?  One avenue to explore concerns photospheric oscillation
properties, which are known to depend on both spectral type and
luminosity class (\#1 and \#3 above), and may depend on rotation and
activity as well (\#2).  If coronal abundance fractionation is
controlled by Alfv\'{e}n and other MHD waves traveling through the
chromosphere, as suggested by \citet{jml04,jml09,jml12}, the
photospheric oscillation characteristics might be expected to be
crucial, since these oscillations are likely sources of many of
the chromospheric waves.

     Oscillation frequencies increase and amplitudes decrease towards
later type stars on the main sequence \citep{hk95,hk11,wjc08},
which have higher $F_{bias}$ values.  This
would imply a potential connection between high $F_{bias}$ and
oscillations with high frequency and/or low amplitude.  Unfortunately,
this hypothetical connection is not consistent with the
$\pi^3$~Ori/Procyon dichotomy.  Oscillation frequencies decrease and
amplitudes increase as stars evolve off the main sequence
\citep{hk95,hk11,wjc08}, so the aforementioned main
sequence oscillation/$F_{bias}$ connection would predict {\em lower}
$F_{bias}$ for Procyon compared with $\pi^3$~Ori.  This is
inconsistent with what we observe, so photospheric oscillation
properties cannot be the sole determinant of coronal abundances.

\subsection{A Possible Theoretical Framework for Resolving the Coronal
  Abundance Problem}

     Turning to theoretical considerations, we outline a
framework for trying to interpret the observed coronal abundance
variations in solar-like stars.  \citet{jml09,jml12} and \citet{cer12}
argue that the Alfv\'{e}n waves giving rise to the
ponderomotive force that generates the solar FIP effect most plausibly
have a coronal origin.  This is because waves in resonance with a
coronal loop having period $\tau = 2L/nv_A$, where $L$ is the coronal
loop length and $v_A$ is the coronal Alfv\'{e}n speed, are required to
produce the observed coronal depletion of He, and the only way to
guarantee such a matching is if processes in the corona (e.g.,
nanoflares) excite normal modes of the loop.  \citet{bew12}
present a ``toy model'' of the inverse FIP effect, where fast mode
waves propagating up through the chromosphere from below are reflected
back down again in the region of the chromosphere where the Alfv\'{e}n
speed is increasing with height. This wave population gives a
ponderomotive force directed {\em downwards}, which competes with the
{\em upwards} ponderomotive force induced by the coronal Alfv\'{e}n
waves propagating downwards and reflecting back up.  We argue
that at the F and G dwarf end of the FBST relation, the positive FIP
effect from coronal Alfv\'{e}n waves dominates, while at the M dwarf
end the inverse FIP effect due to initially upward propagating fast
mode waves dominates, with a transition between the two somewhere in
the K dwarfs.

     Why should this be so?  We discuss first the coronal Alfv\'{e}n
waves.  These are assumed to arise from coronal nanoflare reconnection
events.  Several authors \citep[e.g.,][]{dwl12,hk10,pas99}
have argued that at least some, and maybe
even most, of the magnetic energy released in reconnection should be
converted to kinetic energy or waves in the surrounding plasma.  It is
plausible that such wave generation explains why surveys to find
localized hot plasma as evidence of nanoflare reconnection have
generally been unsuccessful \citep[e.g.,][]{hpw11}.  Instead,
energy goes from magnetic field to waves, and is thus gradually
dissipated as heat throughout the corona, and not quickly and locally
as might have been expected.

     \citet{jfd06} studied the efficiency of electron
acceleration in reconnection with the ambient plasma $\beta$ (the
electron plasma $\beta _e = 8\pi n_ek_{\rm B}T_e/B^2$, to be more
precise).  They found maximum energy input to electrons at $\beta
_e=0$, with reduced electron heating at higher $\beta _e$, or lower
magnetic field.  We suggest that at the left hand side of the FBST
relation, coronal reconnection primarily generates Aflv\'{e}n waves
that end up causing positive FIP fractionation when they reflect from
the chromosphere.  As one moves to the right, to later spectral type,
coronal reconnection puts more energy into electrons, and less into
waves.  Consequently, the positive FIP effect diminishes.

     Conversely, it is also possible that the amplitudes of fast mode
waves penetrating the chromosphere from below and reflecting back down
again may increase toward later spectral types, thus increasing the
inverse FIP effect.  Even though p-mode amplitudes decrease towards
later type stars \citep{hk95,hk11,wjc08}, it is possible that later
type stars could be more efficient at converting the p-mode oscillations
to fast mode MHD waves.  Different regions of the Sun are different in
this respect.  It is known that sunspots are sinks of p-mode energy
\citep[see][and references therein]{dcb95}, most likely through mode
conversion or resonant absorption to Alfv\'{e}n or fast mode waves.
There are many cases of sunspots, active regions, and flares observed
to have reduced FIP effects \citep[e.g.][]{uf90,kjhp94,kjhp06}, although
it is useful to note that an inverse FIP effect has never been
reported in any solar observation.

     An important factor that could act to restrict inverse FIP
effects to later spectral types is the expansion of the magnetic field
within the chromosphere.  In their ``toy model,'' \citet{bew12}
found that the density and magnetic field scale heights, $H_D$ and
$H_B$, were required to satisfy $\left|H_D\right| <
\left|H_B\right|/6$ to yield an inverse FIP effect.  Although more
realistic models are required to refine this relation, we note that
$H_D$ should be smaller for later spectral types due to higher surface
gravities, allowing the relation to be met more readily if $H_B$ does
not also systematically decrease with spectral type.  It is also worth
noting that this relation will be met more readily for extremely
active stars with magnetic fields so crowded that the fields cannot
expand in the chromosphere, which effectively means an increase in
$H_B$.  This could potentially explain the tendency towards high
$F_{bias}$ for extremely active stars (dependence \#2 in
section 3.2).

     The $\pi^3$~Ori/Procyon dichotomy (see section 3.1) remains
difficult to explain within this framework.  Procyon would have larger
$H_D$ than $\pi^3$~Ori, which would bias the star towards lower
$F_{bias}$, not higher (as observed), assuming magnetic field
expansion is the central issue.  If $H_B$ increases by much more than
$H_D$ as one moves from the main sequence to more evolved stars, that
would reverse this behavior, but we know of no physical reason to
expect this to be the case.

     Perhaps Procyon has a fundamentally different coronal heating
mechanism than the stars that follow the FBST relation.  We have
argued in section 2.4 that the commonly observed emission measure
spike at $\log T \simeq 6.6$ would be inconsistent with a Type II
spicule model of coronal heating and mass supply, if interpreted in
terms of a thermal stability argument.  But Procyon does not possess
this peak (see section 3.1), so a spicule-based coronal heating
picture might work for Procyon.  Such a model could also be consistent
with Procyon's lack of a FIP effect, because the spicule flow speed up
through the chromosphere would conceivably be too fast to allow
fractionation to occur.  In this picture, the lower surface gravity of
Procyon would allow upward propagating spicules to reach higher
altitude and presumably higher temperature than on main sequence
stars, potentially allowing the spicules to dominate the rather cool
coronal emission observed from the star.  Drake et al.\ (1995, their
subsection 6.2) previously argued that the absence of FIP effect in
the corona of Procyon is due to the effect of ``unresolved fine
structures'' extending to higher temperatures there than they do on
the Sun.

     These speculations provide a framework for trying to understand
how coronal abundances depend on stellar parameters.  We hope this
discussion can serve to focus attention on the critical observational
and theoretical points required to subject it to further scrutiny.



\acknowledgments

We would like to thank Harry Warren for providing us with the
solar active region EM distribution in Figure~2, and the referee
Manuel G\"{u}del for helpful comments.
Support for this work was provided by NASA through ATP award
NNH11AQ23I and Chandra Award Number GO1-12012Z issued by the Chandra
X-ray Center (CXC).

\clearpage

\begin{deluxetable}{lcccc}
\tabletypesize{\scriptsize}
\tablecaption{{\em Chandra} Line Measurements}
\tablecolumns{5}
\tablewidth{0pt}
\tablehead{
  \colhead{Ion} & \colhead{$\lambda_{rest}$ (\AA)} &
    \colhead{$\log T$} & \colhead{Counts} &
    \colhead{Flux ($10^{-5}$ cm$^{-2}$ s$^{-1}$)}}
\startdata
Si XIII  &   6.648 &6.99 & $ 142.0\pm 30.3$ & $4.14\pm 0.88$ \\
Si XIII  &   6.688 &6.99 &                  &                \\
Si XIII  &   6.740 &6.99 &                  &                \\
Mg XII   &   8.419 &7.11 & $  56.9\pm 20.2$ & $1.97\pm 0.70$ \\
Mg XII   &   8.425 &7.11 &                  &                \\
Mg XI    &   9.169 &6.80 & $ 164.2\pm 27.3$ & $6.63\pm 1.10$ \\
Mg XI    &   9.231 &6.80 &                  &                \\
Mg XI    &   9.314 &6.79 &                  &                \\
Ne X     &  12.132 &6.87 & $ 191.7\pm 23.8$ & $8.89\pm 1.10$ \\
Ne X     &  12.138 &6.87 &                  &                \\
Fe XVII  &  12.264 &6.62 & $ 117.3\pm 21.4$ & $5.47\pm 1.00$ \\
Fe XXI   &  12.285 &6.98 &                  &                \\
Ni XIX   &  12.435 &6.69 & $  74.2\pm 20.1$ & $3.48\pm 0.94$ \\
Ne IX    &  13.447 &6.58 & $ 314.3\pm 36.4$ & $14.41\pm 1.67$ \\
Ne IX    &  13.553 &6.58 &                  &                \\
Ne IX    &  13.699 &6.58 & $  98.2\pm 22.9$ & $4.51\pm 1.05$ \\
Fe XVII  &  13.823 &6.60 & $ 173.2\pm 27.4$ & $7.97\pm 1.26$ \\
Ni XIX   &  14.043 &6.68 & $  39.1\pm 18.9$ & $1.80\pm 0.87$ \\
Ni XIX   &  14.077 &6.68 &                  &                \\
Fe XVIII &  14.203 &6.74 & $ 292.1\pm 33.8$ & $13.47\pm 1.56$ \\
Fe XVIII &  14.208 &6.74 &                  &                \\
Fe XVII  &  15.015 &6.59 & $ 990.1\pm 44.4$ & $44.83\pm 2.01$ \\
Fe XVII  &  15.262 &6.59 & $ 481.6\pm 34.7$ & $21.66\pm 1.56$ \\
O VIII   &  15.176 &6.65 &                  &                \\
Fe XIX   &  15.298 &6.83 &                  &                \\
O VIII   &  16.006 &6.63 & $ 432.2\pm 34.2$ & $19.77\pm 1.56$ \\
Fe XVIII &  16.005 &6.73 &                  &                \\
Fe XVIII &  16.072 &6.73 &                  &                \\
Fe XVII  &  16.778 &6.58 & $ 650.1\pm 36.4$ & $29.20\pm 1.63$ \\
Fe XVII  &  17.053 &6.58 & $1428.3\pm 47.8$ & $76.28\pm 2.55$ \\
Fe XVII  &  17.098 &6.58 &                  &                \\
O VII    &  18.627 &6.34 & $  36.7\pm 17.0$ & $1.93\pm 0.89$ \\
O VIII   &  18.967 &6.59 & $ 624.5\pm 34.8$ & $32.58\pm 1.82$ \\
O VIII   &  18.973 &6.59 &                  &                \\
O VII    &  21.602 &6.32 & $ 152.2\pm 24.5$ & $12.23\pm 1.97$ \\
O VII    &  21.807 &6.32 & $  36.5\pm 19.1$ & $2.98\pm 1.56$ \\
O VII    &  22.101 &6.31 & $ 131.0\pm 23.9$ & $10.75\pm 1.96$ \\
N VII    &  24.779 &6.43 & $  47.5\pm 19.7$ & $3.97\pm 1.65$ \\
N VII    &  24.785 &6.43 &                  &                \\
C VI     &  33.734 &6.24 & $ 114.0\pm 21.2$ & $12.78\pm 2.38$ \\
C VI     &  33.740 &6.24 &                  &                \\
S XIII   &  35.667 &6.43 & $  43.6\pm 18.6$ & $5.09\pm 2.17$ \\
Si XI    &  43.763 &6.25 & $  94.4\pm 20.9$ & $5.31\pm 1.18$ \\
Si XII   &  44.019 &6.44 & $ 112.6\pm 21.4$ & $6.29\pm 1.20$ \\
Si XII   &  44.165 &6.44 & $ 191.1\pm 23.8$ & $10.67\pm 1.33$ \\
Si XII   &  45.521 &6.44 & $  34.6\pm 19.1$ & $1.97\pm 1.09$ \\
Si XII   &  45.691 &6.44 & $  92.7\pm 23.0$ & $5.30\pm 1.31$ \\
Fe XVI   &  46.661 &6.43 & $ 116.7\pm 23.6$ & $6.77\pm 1.37$ \\
Fe XVI   &  46.718 &6.43 &                  &                \\
Si XI    &  49.222 &6.24 & $  55.6\pm 19.8$ & $4.35\pm 1.55$ \\
Fe XVI   &  50.361 &6.43 & $ 151.1\pm 24.4$ & $15.81\pm 2.55$ \\
Fe XVI   &  50.565 &6.43 & $  30.3\pm 16.3$ & $3.60\pm 1.94$ \\
Si X     &  50.524 &6.15 &                  &                \\
Si X     &  50.691 &6.15 & $  37.4\pm 16.6$ & $4.99\pm 2.21$ \\
Si XI    &  52.298 &6.24 & $  14.9\pm 14.1$ & $2.15\pm 2.04$ \\
Fe XVI   &  54.127 &6.43 & $  66.6\pm 20.5$ & $9.06\pm 2.79$ \\
Fe XVI   &  54.710 &6.43 & $ 117.7\pm 22.1$ & $13.67\pm 2.57$ \\
Mg X     &  57.876 &6.22 & $  64.3\pm 19.6$ & $6.05\pm 1.84$ \\
Mg X     &  57.920 &6.22 &                  &                \\
Fe XV    &  59.405 &6.32 & $  66.6\pm 19.4$ & $6.43\pm 1.87$ \\
Fe XVI   &  62.872 &6.42 & $  41.3\pm 16.4$ & $5.71\pm 2.27$ \\
Mg X     &  63.152 &6.21 & $  19.2\pm 13.0$ & $2.84\pm 1.93$ \\
Mg X     &  63.295 &6.21 & $  37.0\pm 16.2$ & $5.70\pm 2.49$ \\
Fe XVI   &  63.711 &6.42 & $  86.9\pm 18.9$ & $16.77\pm 3.65$ \\
Fe XVI   &  66.249 &6.42 & $ 173.2\pm 26.1$ & $26.63\pm 4.01$ \\
Fe XVI   &  66.357 &6.42 &                  &                \\
Fe XV    &  69.682 &6.32 & $  94.3\pm 20.2$ & $9.97\pm 2.14$ \\
Fe XV    &  69.941 &6.32 & $  22.3\pm 14.9$ & $2.38\pm 1.59$ \\
Fe XV    &  69.987 &6.32 &                  &                \\
Fe XV    &  70.054 &6.32 &                  &                \\
Fe XV    &  73.472 &6.32 & $  45.4\pm 17.9$ & $5.39\pm 2.13$ \\
Fe XVI   &  76.497 &6.42 & $  42.2\pm 18.1$ & $5.52\pm 2.37$ \\
Ne VIII  &  88.082 &5.96 & $  50.2\pm 19.7$ & $8.46\pm 3.32$ \\
Ne VIII  &  88.120 &5.96 &                  &                \\
Fe XVIII &  93.923 &6.68 & $ 303.0\pm 36.8$ & $54.48\pm 6.62$ \\
Fe XIX   & 101.550 &6.80 & $  29.8\pm 20.1$ & $6.66\pm 4.49$ \\
Fe XVIII & 103.937 &6.68 & $  82.0\pm 29.1$ & $18.86\pm 6.69$ \\
Fe XIX   & 108.355 &6.79 & $ 133.1\pm 30.5$ & $30.85\pm 7.07$ \\
Fe XX    & 121.845 &6.88 & $  82.9\pm 28.8$ & $23.14\pm 8.04$ \\
Fe XXI   & 128.752 &6.95 & $  31.9\pm 19.9$ & $13.27\pm 8.28$ \\
Fe XX    & 132.840 &6.88 & $  88.2\pm 32.0$ & $36.44\pm 13.22$ \\
Fe XXIII & 132.906 &7.12 &                  &                \\
Fe XXII  & 135.755 &7.03 & $  38.9\pm 24.1$ & $15.73\pm 9.75$ \\
Fe IX    & 171.073 &5.95 & $  33.0\pm 22.6$ & $46.92\pm 32.13$ \\
Fe X     & 174.534 &6.04 & $  33.9\pm 22.5$ & $125.82\pm 83.51$ \\
\enddata
\end{deluxetable}

\clearpage

\begin{deluxetable}{cccc}
\tabletypesize{\normalsize}
\tablecaption{Elemental Abundances}
\tablecolumns{4}
\tablewidth{0pt}
\tablehead{
  \colhead{} & \colhead{Stellar} & \colhead{Stellar} & \colhead{Solar} \\
  \colhead{Abundance} & \colhead{Corona} &
    \colhead{Photosphere\tablenotemark{a}} &
    \colhead{Photosphere\tablenotemark{b}}}
\startdata
${\rm Fe/Fe_{\odot}}$  & 1.5                & 1.0    & 1.0 \\
$\log {\rm [Fe/H]}$  & $-4.32$         & $-4.50$ & $-4.50$ \\
$\log {\rm [C/Fe]}$  & $0.70_{-0.11}^{+0.14}$ & 1.12 &  0.93 \\
$\log {\rm [N/Fe]}$  & $-0.03_{-0.30}^{+0.18}$ &(0.49)&  0.33 \\
$\log {\rm [O/Fe]}$  & $0.73_{-0.04}^{+0.03}$ & 1.35 &  1.19 \\
$\log {\rm [Ne/Fe]}$ & $0.30_{-0.06}^{+0.05}$ &(0.96)&  0.43 \\
$\log {\rm [Mg/Fe]}$ & $0.08_{-0.06}^{+0.12}$ & 0.08 &  0.10 \\
$\log {\rm [Si/Fe]}$ & $0.16_{-0.07}^{+0.06}$ & 0.07 &  0.01 \\
$\log {\rm [S/Fe]}$  &$-0.13_{-0.64}^{+0.16}$ &($-0.22$)&  $-0.38$\\
$\log {\rm [Ni/Fe]}$ &$-1.21_{-0.25}^{+0.12}$ & $-1.33$ &  $-1.28$\\
\enddata
\tablenotetext{a}{From Allende Prieto et al.\ (2004), except for values in
  parentheses, which are assumed rather than directly measured (see text).}
\tablenotetext{b}{From Asplund et al.\ (2009).}
\end{deluxetable}

\clearpage

\begin{figure}[t]
\plotfiddle{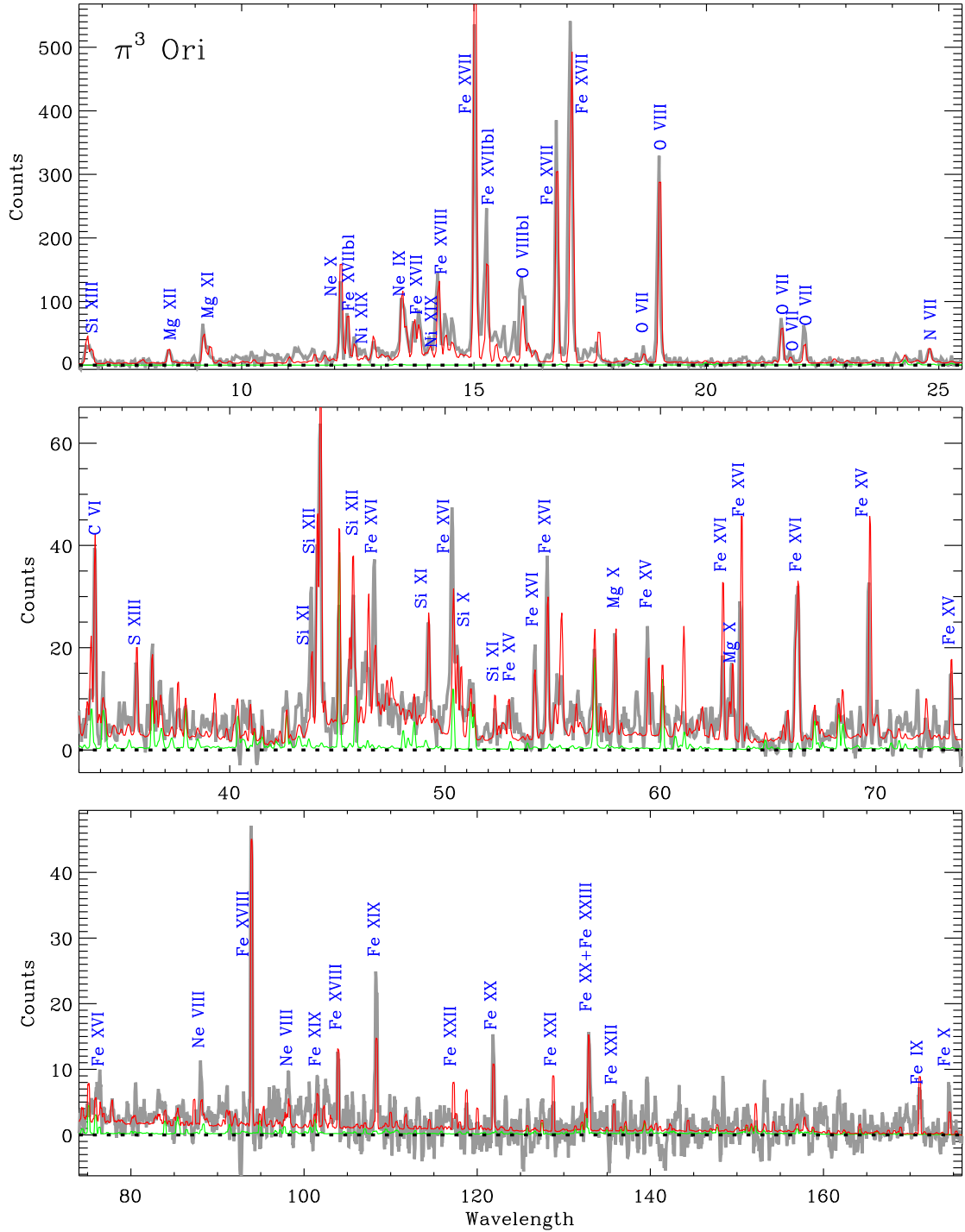}{7.0in}{0}{90}{90}{-300}{-90}
\caption{The {\em Chandra} LETGS spectrum of $\pi^3$~Ori, rebinned by
  a factor of 3 to improve S/N.  For wavelengths above 35~\AA, the
  spectrum is also smoothed for the sake of appearance.  The red line
  is a synthetic spectrum computed using the emission measure
  distribution in Figure~2, and the green lines indicate the
  contributions of higher spectral orders (2--5) to the model
  spectrum.}
\end{figure}

\clearpage

\begin{figure}[t]
\plotfiddle{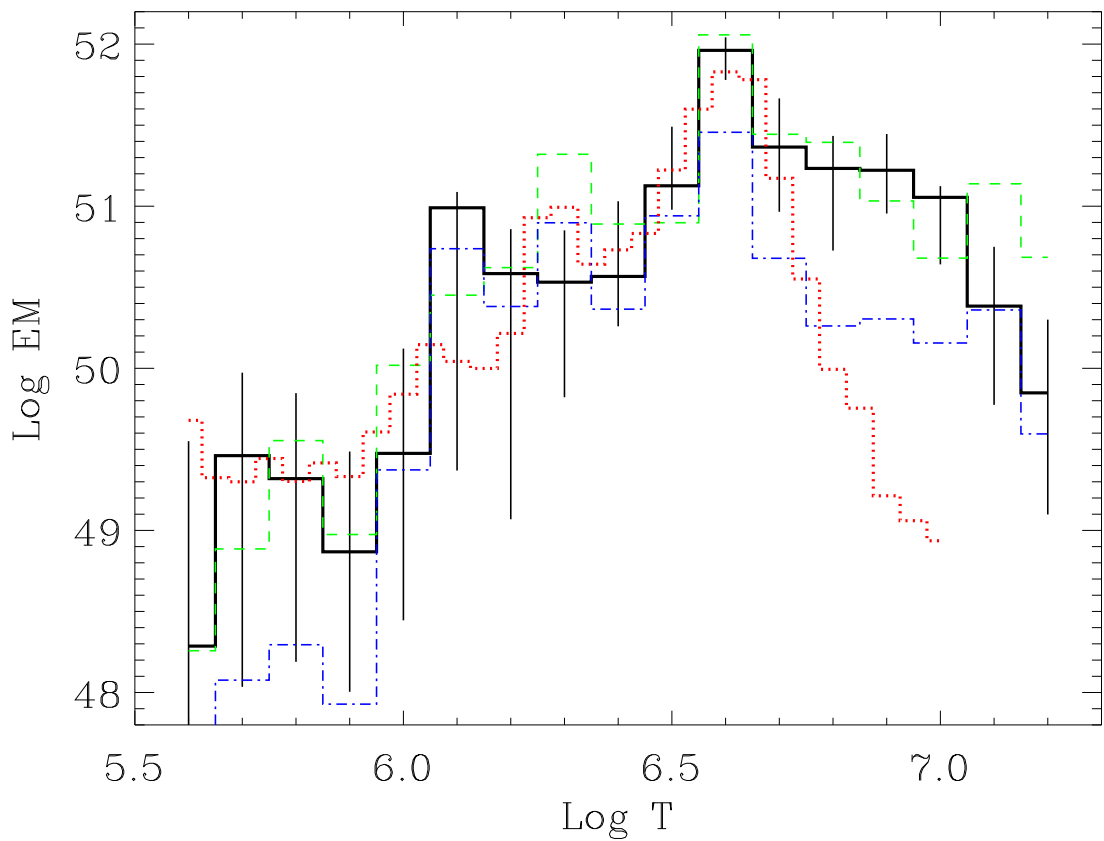}{3.0in}{90}{70}{70}{260}{-50}
\caption{Coronal emission measure distribution derived for
  $\pi^3$~Ori, with 90\% confidence error bars.  This is compared with
  distributions for $\epsilon$~Eri (dot-dashed line) and $\xi$~Boo~A
  (dashed line), which were also measured from {\em Chandra}/LETGS data
  (Wood \& Linsky 2006, 2010).  Also shown for comparison is an
  average emission measure distribution for solar
  active regions (dotted line) studied by Warren et al.\ (2012).  The
  solar distribution is normalized so that its peak matches that of
  $\pi^3$~Ori.}
\end{figure}

\clearpage

\begin{figure}[t]
\plotfiddle{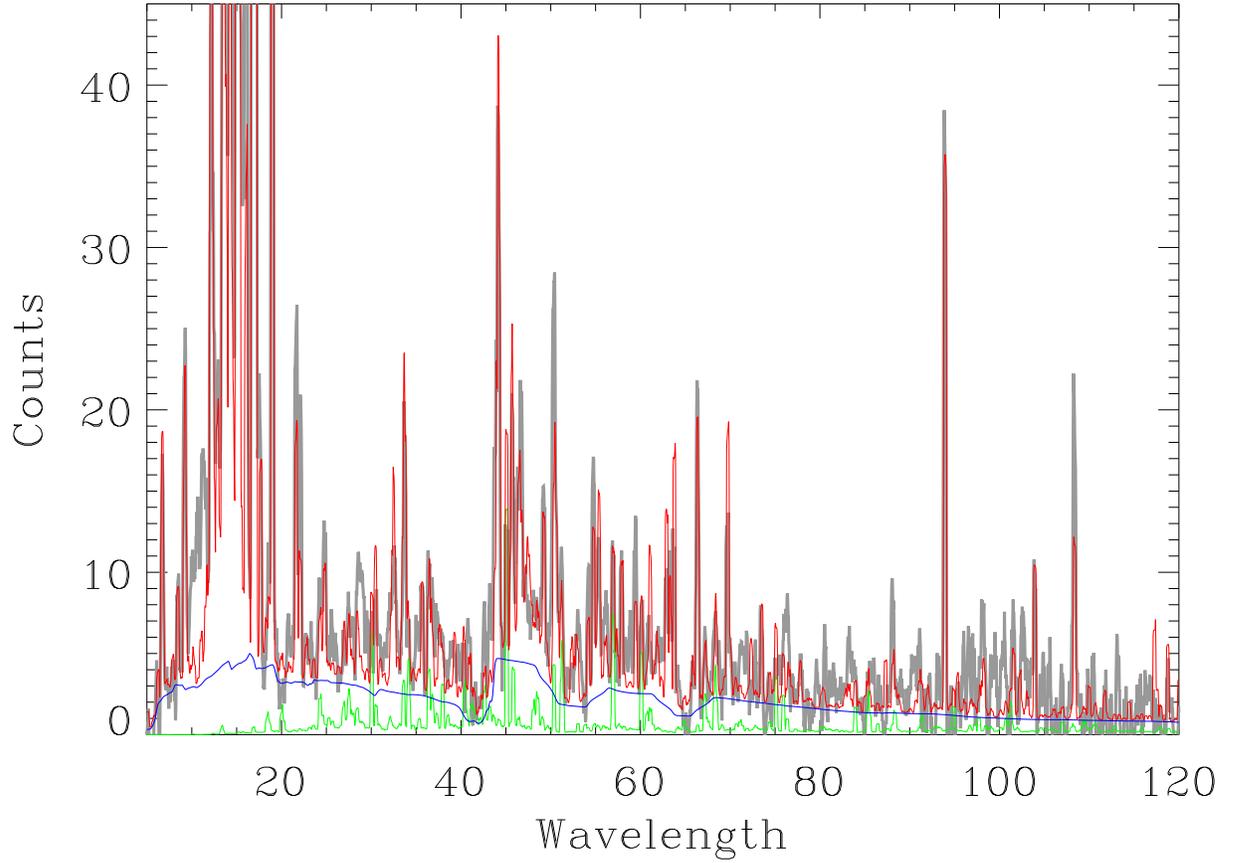}{3.3in}{0}{60}{60}{-210}{-60}
\caption{A synthetic line-plus-continuum spectrum (red line) showing
  the best fit to the highly smoothed {\em Chandra} $\pi^3$~Ori
  spectrum (gray line).  The continuum contribution (blue line) to the
  synthetic spectrum assumes an absolute Fe abundance of $\log {\rm
    [Fe/H]}=-4.32$.  The green line shows the contributions of higher
  spectral orders (2-5) to the total spectrum.}
\end{figure}

\clearpage

\begin{figure}[t]
\plotfiddle{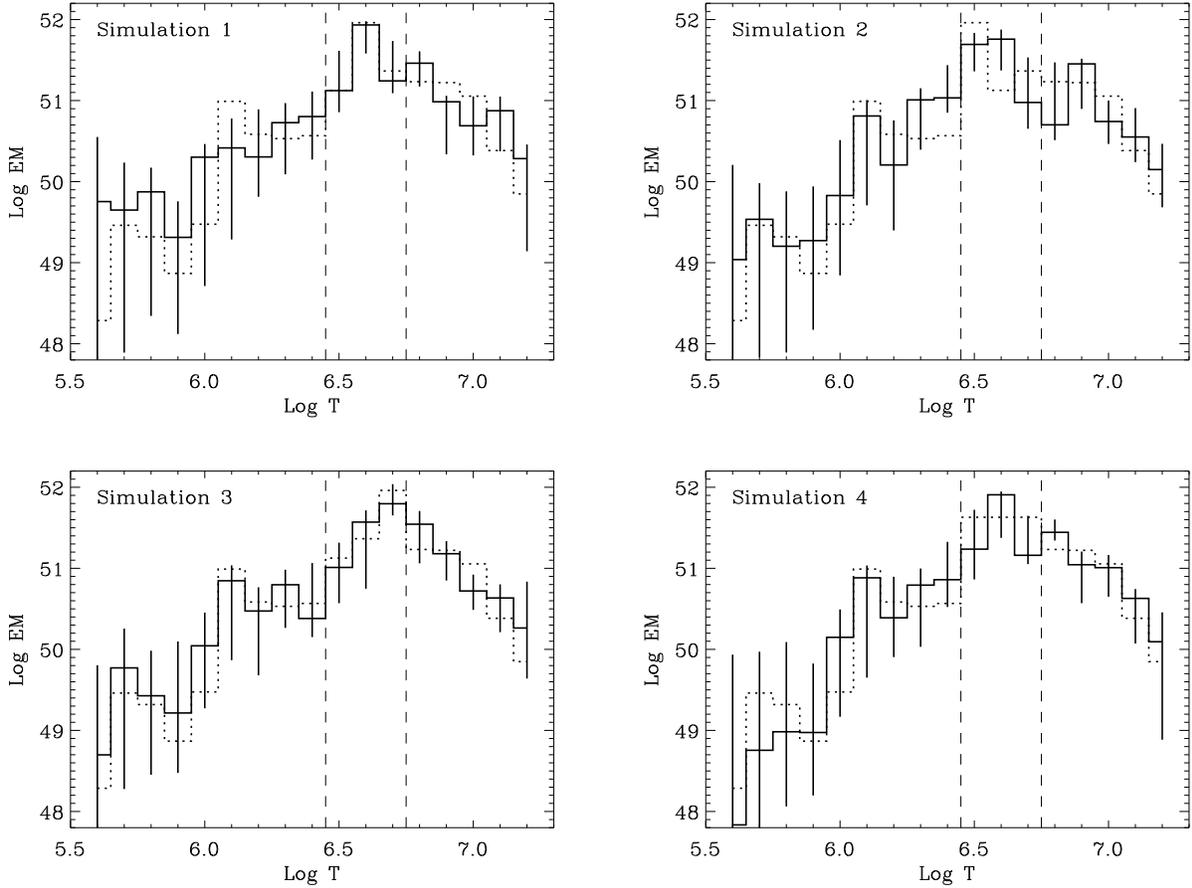}{3.0in}{90}{70}{70}{270}{-50}
\caption{Illustration of four simulations used to assess the
  reliability of EM computations by PINTofALE.  Dotted lines in each
  panels are emission measure distributions used to compute line
  fluxes, which are then fed into PINTofALE with the same relative
  uncertainties as the lines measured for $\pi^3$~Ori.  Solid lines
  show the resulting EM distributions, with 90\% confidence intervals.
  In all cases, these agree reasonably well with the actual
  distribution.  For Simulation~1, the initial seed distribution is
  the $\pi^3$~Ori distribution from Fig.~2.  This distribution is also
  used for the other three simulations, but it is modified in the
  $\log T=6.5-6.7$ range (marked by dashed lines) to explore PINTofALE's
  ability to reproduce EM peaks at these temperatures.}
\end{figure}
\clearpage

\begin{figure}[t]
\plotfiddle{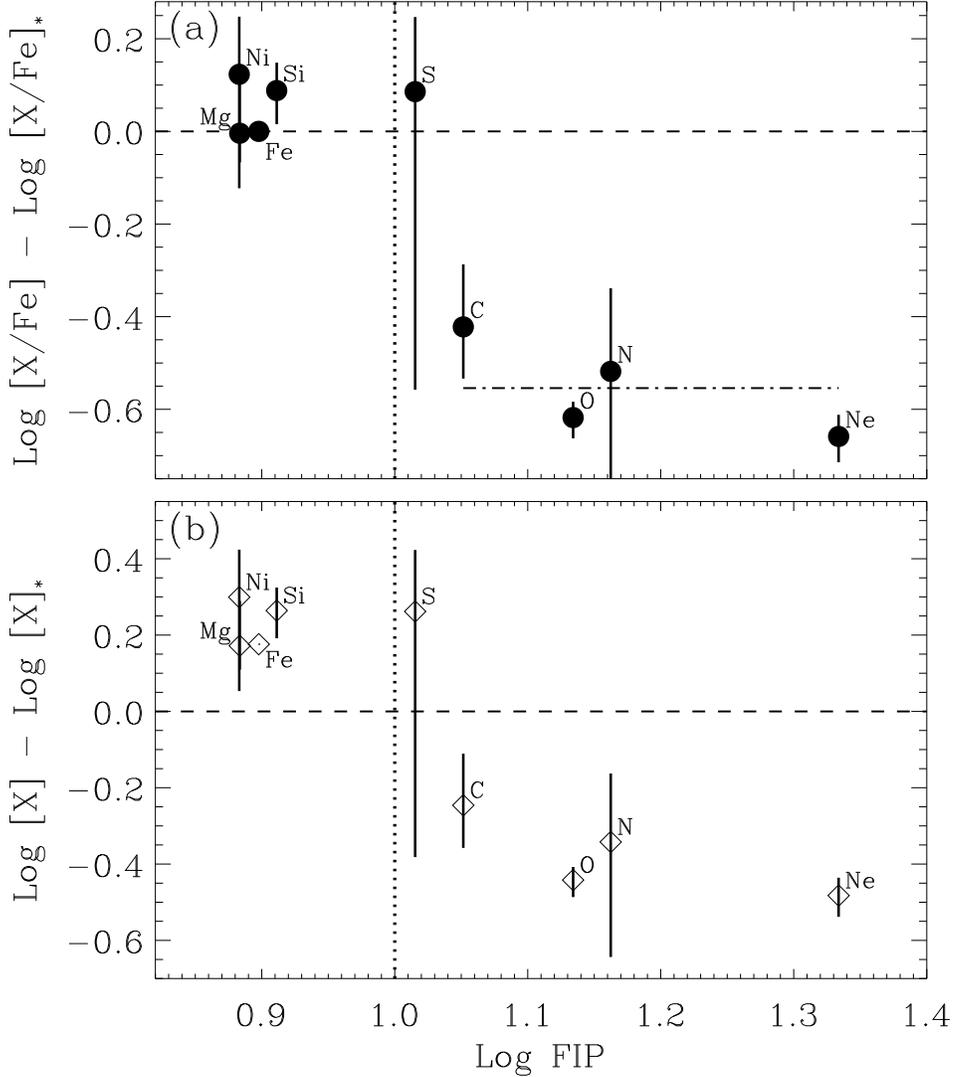}{5.5in}{90}{50}{50}{90}{-40}
\caption{(a) The coronal abundance ratios of elements relative to Fe
  ([X/Fe]) are normalized to the photospheric abundance ratios
  ([X/Fe]$_*$) and plotted versus first ionization potential (i.e.,
  FIP), in eV.  Error bars are 90\% confidence intervals derived from
  the emission measure analysis.  The vertical dotted line separates
  low-FIP elements from high-FIP elements.  The horizontal dot-dashed
  line is the average relative abundance of the four principal
  high-FIP elements (C, N, O, Ne), which is the $F_{bias}$ quantity
  used in Fig.~6.  (b) Absolute coronal abundances of $\pi^3$~Ori
  relative to the photosphere, based on the $\log {\rm [Fe/H]}=-4.32$
  measurement from the line-to-continuum ratio analysis (see Fig.~3).}
\end{figure}

\clearpage

\begin{figure}[t]
\plotfiddle{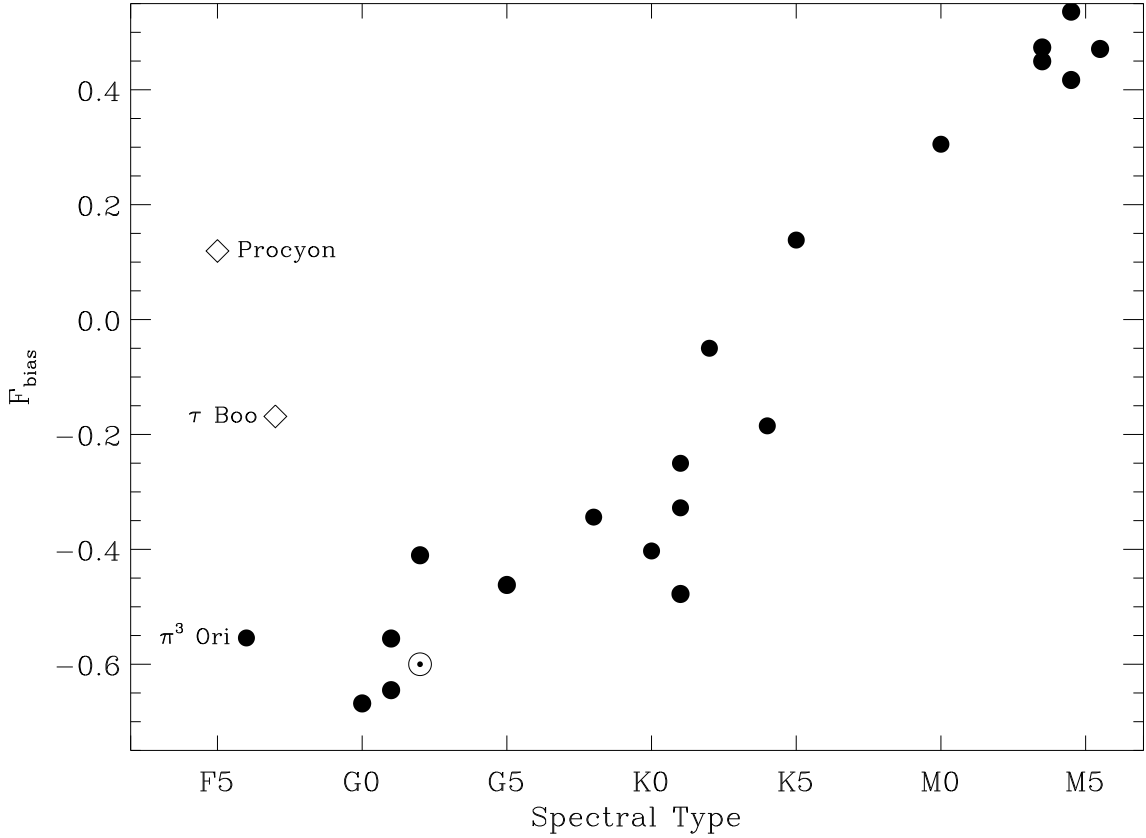}{3.0in}{90}{40}{40}{50}{-30}
\caption{A plot of FIP bias ($F_{bias}$) versus spectral type for the
  sample of main sequence stars from Wood et al.\ (2012), but with the
  addition of our new $\pi^3$~Ori measurement.  Measurements are also
  shown for two additional F stars, Procyon and $\tau$~Boo, which are
  clearly inconsistent with $\pi^3$~Ori and the general
  FIP-bias/spectral-type (FBST) relation.}
\end{figure}

\end{document}